\centerline{\bf A short table of generating functions and related formulas}   
\bigskip
 \centerline{Robert M. Ziff}        
 \centerline{Michigan Center for Theoretical Physics and Department of Chemical Engineering}        
 \centerline{University of Michigan, Ann Arbor, MI 48109-2136 USA}        
  \centerline{rziff@umich.edu}        
\bigskip

A table of sums useful for generating function applications (discrete Laplace transforms or $z$-transforms).  
Related definitions and formulas (including Lagrange's expansion), and reference to formulas
in Abramowitz and Stegun {\sl Handbook of Mathematical Functions} are given.

Bold-face equation numbers refer to formulas in Abramowitz and Stegun.  Many of
these can be found or verified using Mathematica.


\bigskip
\medskip                                            
                                
\bigskip
            \centerline{\bf Notation and Definitions of Functions}  
\bigskip
In general, $z$, $r$, and $s$ are complex numbers (although some formulas  may be valid for real values only),
$x$ a real number, $n$ and $m$ are integers, ${\cal R}z$ means the real part of $z$.
$$\leftline{\indent$\displaystyle
(z)_n = { \Gamma(z+n) \over \Gamma(z) } = z (z+1) \ldots (z + n - 1) \ , \qquad (z)_0 = 0                   \hfill            {\bf 6.1.22}  $}$$
   
$$\leftline{\indent$\displaystyle
{r \choose n} = {r (r - 1) (r -2) \ldots (r-n+1) \over n!},\qquad {m \choose n} = {m! \over n! (m-n)!} \qquad (m \ge n) \hfill   {\bf 24.1.1}  $}$$
  
  \centerline{\bf Sums and Functions}   
 \bigskip                               
\noindent Binomial formula (in general, valid for $|z| < 1$):                                                         
$$\leftline{\indent$\displaystyle
 \sum_{n=0}^\infty {r \choose n} z^n = (1+z)^r                       \hfill              {\bf 3.6.8} $}$$ 

$$\leftline{\indent$\displaystyle                                                         
 \sum_{n=0}^\infty {n+s-1 \choose n} z^n   
 = \sum_{n=0}^\infty {\Gamma(n+s) \over \Gamma(s)} {z^n \over n!}   
 = \sum_{n=0}^\infty (s)_n {z^n \over n!}  
      = (1-z)^{-s}                                                             \hfill             {\bf 3.6.9} $}$$

$$\leftline{\indent$\displaystyle
 \sum_{n=0}^\infty z^n = {1 \over 1-z}                                       \hfill        {\bf 3.6.10} $}$$

$$\leftline{\indent$\displaystyle
 \sum_{n=0}^\infty n z^n =  z {d \over dz}  \sum_{n=0}^\infty z^n  = z {d \over dz}  {1 \over 1-z} = {z \over (1-z)^2}      \hfill         $}$$ 
   
$$\leftline{\indent$\displaystyle
 \sum_{n=0}^\infty n^2 z^n = \left( z {d \over dz} \right)^2  {1 \over 1-z} 
                           = z {d \over dz} {z \over (1-z)^2} =  {z+z^2 \over (1-z)^3}                           \hfill        $}$$ 
                                                   
$$\leftline{\indent$\displaystyle
 \sum_{n=0}^\infty n^3 z^n = {z+4z^2+z^3 \over (1-z)^4}                     \hfill        $}$$         

$$\leftline{\indent$\displaystyle
 \sum_{n=0}^\infty n^4 z^n = {z+11z^2+11z^3+z^4 \over (1-z)^5}                     \hfill        $}$$         

$$\leftline{\indent$\displaystyle
 \sum_{n=0}^\infty n(n-1) z^n = 2 \sum_{n=0}^\infty {n \choose 2} z^n 
     = z^2 \left( {d \over dz} \right)^2 \sum_{n=0}^\infty z^n = {2 z^2 \over (1-z)^3 }      \hfill    $}$$ 

$$\leftline{\indent$\displaystyle
 {1 \over i!}\sum_{n=0}^\infty [n(n-1)(n-2)\ldots(n-i+1)] z^n
                     = \sum_{n=0}^\infty {n \choose i} z^n = { z^i \over (1-z)^{i+1} }      \hfill    $}$$

$$\leftline{\indent$\displaystyle
 \sum_{n=0}^\infty (n+1) z^n
                     = { 1 \over (1-z)^2 }      \hfill    $}$$ 

$$\leftline{\indent$\displaystyle
 {1 \over 2}\sum_{n=0}^\infty (n+1)(n+2) z^n
                     = { 1 \over (1-z)^3 }      \hfill    $}$$

$$\leftline{\indent$\displaystyle
 {1 \over i!}\sum_{n=0}^\infty [(n+1)(n+2)\ldots(n+i)] z^n
                     = \sum_{n=0}^\infty {n + i\choose i} z^n = { 1 \over (1-z)^{i+1} }      \hfill    $}$$

$$\leftline{\indent$\displaystyle
 \sum_{n=0}^\infty  {2n \choose n} { z^n \over 2^{2n} }  
   = \sum_{n=0}^\infty \left(\textstyle{1 \over 2}\right)_n {z^n \over n!}  =  (1-z)^{-1/2}   $}$$

$$\leftline{\indent$\displaystyle
 \sum_{n=0}^\infty {2n \choose n} {z^n \over 2^{2n}(n+1)}  = { 2[1-(1-z)^{1/2}] \over z }   $}$$   

$$\leftline{\indent$\displaystyle
 \sum_{n=0}^\infty {2n \choose n} {z^n \over 2^{2n}(n+1)(n+2)}  = {4[ (1-z)^{3/2} - 1 + 3z/2] \over 3 z^2 }   $}$$   
            
\medskip                                                                                                      
\noindent log, inverse trigonometric functions:   

$$\leftline{\indent$\displaystyle
 \sum_{n=1}^\infty {z^n \over n} = \int_0^z\left( \sum_{n=1}^\infty t^{n-1} \right) dt = \int_0^z {dt \over 1-t}  = - \ln (1-z)    \hfill  {\bf 4.1.24} $}$$ 
 
$$\leftline{\indent$\displaystyle
 \sum_{n=1}^\infty {t^n \over n^2} = \int_0^z\left( \sum_{n=1}^\infty {t^{n-1} \over n} \right) dz = - \int_0^z {\ln(1-t) \over t} dt 
              \hbox{   = Euler's dilogarithm = $g_2(z)$ }                                                           \hfill  {\bf 27.7.1} $}$$   
 
$$\leftline{\indent$\displaystyle                             
 \sum_{n=0}^\infty { (\pm 1)^n  z^{2n+1}  \over 2n+1}      
       =  \int_0^z {dt \over 1 \mp t^2} 
       =  \cases{\hbox{arc}\tanh z = (1/2)    \ln[(1+z)/(1-z)] & $(-)$ \cr
                 \hbox{arc}\tan z & $(+)$  \cr                                         }                   \hfill{\bf4.6.22,33\atop4.4.42} $}$$
 
 $$\leftline{\indent$\displaystyle
 \sum_{n=0}^\infty  {2n \choose n} { (\pm 1)^n z^{2n+1}  \over 2^{2n}(2n+1)}   
        =  \int_0^z {dt \over (1 \mp t^2)^{1/2}} 
        =  \cases{ \hbox{arc}\sin z & $(+)$  \cr \hbox{arc}\sinh z = \ln[z+(z^2+1)^{1/2}] & $(-)$ \cr}       \hfill{\bf4.4.40\atop4.6.31} $}$$


$$\leftline{\indent$\displaystyle
 \sum_{n=0}^\infty { (\pm 1)^n 2^{2n} (n!)^2 z^{2n}  \over (n+1)(2 n + 1)!}   
           =  \cases{ \left(\displaystyle {\hbox{arc} \sin z \over z} \right)^2   & $(+)$ \cr
                  \left(\displaystyle {\hbox{arc}\sinh z \over z} \right)^2   & $(-)$ \cr }             $}$$

\medskip
\noindent Exponential, trigonometric, hyperbolic functions: 

$$\leftline{\indent$\displaystyle
 \sum_{n=0}^\infty { z^n \over n! }  =  e^z                                                                    \hfill {\bf 4.2.1} $}$$

$$\leftline{\indent$\displaystyle
 \sum_{n=0}^\infty {(\pm 1)^n z^{2n+1}  \over (2n+1)! }  = 
        \cases{ \sinh z = (e^z - e^{-z})/2 & $(+)$  \cr \sin z = (e^{iz} - e^{-iz})/2i & $(-)$ \cr}          \hfill{\bf4.5.62\atop4.3.65} $}$$

$$\leftline{\indent$\displaystyle
 \sum_{n=0}^\infty {(\pm 1)^n z^{2n}  \over (2n)! }  = 
        \cases{ \cosh z = (e^z + e^{-z})/2& $(+)$  \cr \cos z  = (e^{iz} + e^{-iz})/2& $(-)$ \cr}            \hfill{\bf4.5.63\atop4.3.66} $}$$

\medskip \noindent Exponential and Fresnel Integrals       

$$\leftline{\indent$\displaystyle
 \sum_{n=1}^\infty {(\pm 1)^n z^{n}  \over n\, n! }   
      = \int_0^z \left( \sum_{n=1}^\infty {(\pm 1)^n t^{n-1} \over n!} \right) dt
      = \int_0^z  {e^{\pm t} - 1 \over t} dt
       =  \cases{ \hbox{Ei}(z) - \gamma -\ln z & $(+)$  \cr
                  -E_1(z) - \gamma - \ln z      & $(-)$  \cr}                                                 \hfill{\bf5.1.10\atop5.1.11} $}$$

$$\leftline{\indent$\displaystyle
 \sum_{n=0}^\infty {(\pm 1)^n z^{2n+1}  \over (2n+1)(2n+1)! }  
        = \int_0^z { {\sinh \choose \sin}t \over t} dt
        = \cases{ \hbox{Shi}(z)     & $(+)$     \cr            
                  \hbox{Si}(z)      & $(-)$     \cr  }                                                   \hfill{\bf 5.2.3,17\atop5.2.1,19} $}$$

$$\leftline{\indent$\displaystyle
      \sum_{n=1}^\infty {(\pm 1)^n z^{2n}  \over 2n(2n)! } 
        = \int_0^z  { {\cosh \choose \cos}t - 1 \over t} dt
         = \cases{ \hbox{Chi}(z) - \gamma -\ln z & $(+)$  \cr 
                   \hbox{Ci}(z) - \gamma - \ln z  & $(-)$ \cr}                                           \hfill{\bf 5.2.4,18\atop5.2.2,16} $}$$

\medskip\noindent
Incomplete gamma function and error function

$$\leftline{\indent$\displaystyle
 e^{-z} \sum_{n=0}^\infty { z^{n}  \over \Gamma(a+n+1) }  = 
 {1 \over \Gamma(a)} \sum_{n=0}^\infty { (-1)^n z^{n}  \over (a+n) n! }  =  \gamma^*(a,z)   \hfill {\bf 6.5.29} $}$$    

$$\leftline{\indent$\displaystyle
 e^{-z^2} \sum_{n=0}^\infty { 2^{2n} n!  z^{2n+1}\over (2n+1)! }   
   =  \sum_{n=0}^\infty { (-1)^n z^{2n+1}  \over (2n+1) n! } 
    =  {\sqrt{\pi} \over 2} \hbox{erf } z =  {\sqrt{\pi} \over 2} z \gamma^*(\textstyle{1 \over 2}, z^2)                 \hfill \bf{7.1.5} $}$$    

\medskip\noindent
Bessel function

$$\leftline{\indent$\displaystyle
 \sum_{n=0}^\infty { (\pm 1)^n  z^{2n}  \over 2^{2n} n! \Gamma(\nu+n+1) }  = 
        \cases{  (2/z)^{\nu} I_\nu(z) & $(+)$  \cr  (2/z)^{\nu} J_\nu(z) & $(-)$ \cr}                             \hfill       {\bf 9.1.10} $}$$

$$\leftline{\indent$\displaystyle
 \sum_{n=0}^\infty { (\pm 1)^n  z^{2n}  \over 2^{2n} n! n! }  = 
        \cases{ I_0(z) & $(+)$  \cr  J_0(z) & $(-)$ \cr}                                                            \hfill      {\bf 9.1.12} $}$$

\medskip\noindent
Elliptic integrals

$$\leftline{\indent$\displaystyle
 \sum_{n=0}^\infty   {2n \choose n}^2 {  z^{n} \over 2^{4n}  }   
        = {2\over\pi} \int_0^{\pi/2} (1-z \sin^2 \theta)^{1/2} d\theta 
        = 2 K(z) /\pi                                                                                                 \hfill   {\bf 17.3.1,11} $}$$

$$\leftline{\indent$\displaystyle
 \sum_{n=1}^\infty   {2n \choose n}^2 {  z^{n} \over 2^{4n} (2n-1)  }   
        = 1-{2\over\pi} \int_0^{\pi/2} (1-z \sin^2 \theta)^{-1/2} d\theta 
        = 1-2 E(z) /\pi                                                                                                 \hfill   {\bf 17.3.2,12} $}$$

\medskip\noindent
Bernouilli functions and numbers [$B_n(0) = B_n, 2^n E_n(\textstyle{1\over 2}) = E_n$ -- {\it not} the exponential-integral function].

$$\leftline{\indent$\displaystyle
 \sum_{n=0}^\infty { B_n(a)  z^{n} \over n! }  =  {z e^{az} \over e^z - 1 }  \hfill |z| < 2 \pi                 \hfill                {\bf 23.1.1} $}$$

$$\leftline{\indent$\displaystyle
 \sum_{n=0}^\infty { E_n(a)  z^{n} \over n! }  =  {2 e^{az} \over e^z + 1 }  \hfill |z| <  \pi                   \hfill               {\bf 23.1.1} $}$$     

\medskip\noindent
Generalized zeta function (Bose functions) (see below)
$$\leftline{\indent$\displaystyle
 \sum_{n=1}^\infty {z^n \over n^s} = g_s(z)                                                                             $}$$     
                        
\bigskip   
                                                       
\medskip\noindent
Definitions of special functions in terms of integrals:

$$\leftline{\indent$\displaystyle
\Gamma(z) = \int_0^\infty{t^{z-1} e^{-t} dt}             \qquad  ({\cal R} z > 0)                                                  \hfill    {\bf 6.1.1}  $}$$
 
$\Gamma(z+1)=z\Gamma(z), \Gamma(n) = n-1!, \Gamma(\textstyle{1 \over 2}) = \sqrt{\pi}$
                                                                                                                 
$$\leftline{\indent$\displaystyle
\gamma^*(a,x) = {x^{-a} \over \Gamma(a) }\int_0^x{t^{a-1} e^{-t}  dt}   = x^{-a}[1-\Gamma(a,x)/\Gamma(a)]                         \hfill     {\bf 6.5.4}  $}$$

$$\leftline{\indent$\displaystyle
\Gamma(a,x) = \int_x^\infty{t^{a-1} e^{-t}  dt}                                               \hfill             {\bf 6.5.2}  $}$$

$$\leftline{\indent$\displaystyle
E_1(z) =  \int_z^\infty {{ e^{-t} \over t}  dt}    \qquad (|\hbox{arg } z| < \pi)                 \hfill        {\bf 5.1.1}  $}$$

$$\leftline{\indent$\displaystyle
E_n(z) =  \int_1^\infty {{ e^{-zt} \over t^n}  dt}  = z^{n-1} \Gamma(1-n,z) \qquad  (n = 0, 1, 2, ... ; {\cal R} z > 0)  \hfill           {\bf 5.1.1}  $}$$
  
$$\leftline{\indent$\displaystyle
\hbox{Ei}(x) =  - \lim_{\epsilon \to 0} \left( \int_{-x}^{-\epsilon} { { e^{-t} \over t}  dt}
           + \int_\epsilon^\infty { { e^{-t} \over t} dt } \right)   \qquad (x > 0)                              \hfill              {\bf 5.1.2}  $}$$
 
$$\leftline{\indent$\displaystyle
\hbox{erf } x =  { 2 \over \sqrt{\pi} }  \int_0^x e^{-t^2}  dt                                                      \hfill             {\bf 7.1.1}  $}$$
 
$$\leftline{\indent$\displaystyle
J_{\nu}(z) = { ( z/2)^\nu \over \pi^{1/2} \Gamma(\nu+1/2) }\int_0^\pi \cos(z \cos \theta) \sin^{2\nu} \theta d\theta      \hfill             {\bf 9.1.20}  $}$$

$$\leftline{\indent$\displaystyle
I_{\nu}(z) = {  (z/2)^\nu \over \pi^{1/2} \Gamma(\nu+1/2) }\int_0^\pi e^{\pm z \cos \theta} \sin^{2\nu} \theta d\theta     \hfill            {\bf 9.6.18}  $}$$

$$\leftline{\indent$\displaystyle
\zeta(s) = {1 \over \Gamma(s)}  \int_0^\infty{t^{s-1} \over e^{t}-1}  dt \qquad {\cal R}s>1                                  \hfill            {\bf 23.2.7}  $}$$

$$\leftline{\indent$\displaystyle
g_s(z) = {1 \over \Gamma(s) } \int_0^\infty {t^{s-1} dt \over z^{-1}e^t - 1 }                                                                 \hfill    $}$$

\medskip\noindent
Generalized zeta functions, expansion for small $\alpha = -\ln z \approx 1-z$:
$$\leftline{\indent$\displaystyle
g_s(z) = \cases{
     \displaystyle{ \Gamma(1-s)\alpha^{n-1} + \sum_{k=0}^\infty  {\zeta(s-k) (-\alpha)^k \over k!}  }  & $s \ne 1, 2, 3, \dots $  \cr 
     \displaystyle{ { (-\alpha)^{s-1} \over (s-1)! } \left[ -\ln \alpha + \sum_{m=1}^{s-1} {1 \over m}  \right]
               + \sum_{\textstyle{k=0 \atop k \ne s-1}}^\infty  {\zeta(s-k) (-\alpha)^k \over k!}     }  & $s = 1, 2, 3, \dots $ \cr }                         \hfill    $}$$

\medskip\noindent                                                                                         
Riemann zeta function  [$\zeta(0) = -1/2, \zeta(1) = \infty, \zeta(2) = \pi^2/6$]
$$\leftline{\indent$\displaystyle
\zeta(s) = \sum_{n=1}^\infty {1 \over n^s}   \qquad   {\cal R}s > 1                            \hfill   \bf{23.2.1}    $}$$  

\medskip\noindent                                                                                         
Euler's gamma constant:

$$\leftline{\indent$\displaystyle
\gamma =   \lim_{m \to \infty} \left[ \sum_{n=1}^m {1\over n} - \ln m \right] = 0.57721\,56649\,01532\,86060\ldots                \hfill    {\bf 6.1.3}  $}$$

\medskip\noindent
Taylor's expansion of $f(z)$ about $z_0=0$: 

$$\leftline{\indent$\displaystyle
f(z) = \sum_{n=0}^\infty {z^n \over n!} \left[ {d^{n} \over dz^{n} } f(z)  \right]_{z=0}         \hfill   {\bf 3.6.4}  $}$$                           
                       
\bigskip
\centerline{\bf Lagrange's Expansion}   
\medskip\noindent Lagrange's expansion, where $z=f(x), z_0=f(x_0), f'(x_0)\ne0$:
$$\leftline{\indent$\displaystyle
x = f^{-1}(z) = x_0+\sum_{n=1}^\infty {(z-z_0)^n \over n!} \left[ {d^{n-1} \over dx^{n-1} } \left\{ {x-x_0 \over f(x)-z_0} \right\}^n \right]_{x=x_0}      \hfill {\bf 3.6.6}  $}$$                           
                                                                    
\noindent For $g(x)$ any infinitely differentiable function, 
$$\leftline{\indent$\displaystyle
g(x) = g(0) + \sum_{n=1}^\infty {(z-z_0)^n \over n!} \left[ {d^{n-1} \over dx^{n-1} } \left( g'(x)  \left\{ {x-x_0 \over f(x)-z_0} \right\}^n \right) \right]_{x=x_0}    \hfill   {\bf 3.6.7}  $}$$                           

\medskip\noindent
For special case of $x_0 = 0, z_0=f(x_0)=0$:

\medskip\noindent
$$\leftline{\indent$\displaystyle
x =  \sum_{n=1}^\infty {z^n \over n!} \left[ {d^{n-1} \over dx^{n-1} } \left\{ {x \over f(x)} \right\}^n \right]_{x=0}         \hfill   {\bf 3.6.6}  $}$$                           
             
Examples:  (Useful for polymerization, percolation on the Bethe lattice):

$$\leftline{\indent$\displaystyle
1 + r \sum_{n=1}^\infty {(2n+r-1)! z^n \over(n+r)! n!} = 
1 + r \sum_{n=1}^\infty (n+r-1)_{n-1} {z^n\over n!}= 
r \sum_{n=0}^\infty  {2n+r \choose n} {z^n\over 2n + r} = 
\left[ 1-(1-4z)^{1/2} \over 2z \right]^r   $}$$             

related:
$$\leftline{\indent$\displaystyle
\sum_{n=0}^\infty {2n+r \choose n} z^n = 
{1 \over \sqrt{1 - 4 z}}  \left[ 1-(1-4z)^{1/2} \over 2z \right]^r  $}$$                           
                                   
Taking $z = x e^{-x}:$
$$\leftline{\indent$\displaystyle
 \sum_{n=1}^\infty {n^{n-1}\over  n!}(x e^{-x})^n  
= x   $}$$                           
           
differentiating/integrating w.r.t. $x$:                                  
$$\leftline{\indent$\displaystyle
 \sum_{n=1}^\infty {n^{n}\over n!}(x e^{-x})^n  
= {x \over 1-x}   $}$$                           
           
$$\leftline{\indent$\displaystyle
 \sum_{n=1}^\infty {n^{n+1}\over  n!}(x e^{-x})^n  
= {x  \over (1-x)^3}   $}$$                           
           
$$\leftline{\indent$\displaystyle
 \sum_{n=1}^\infty {n^{n+2}\over  n!}(x e^{-x})^n  
= {x (1+2x) \over (1-x)^5}   $}$$                           
           
$$\leftline{\indent$\displaystyle
 \sum_{n=1}^\infty {n^{n-2}\over  n!}(x e^{-x})^n  
= x - {x^2 \over 2}   $}$$                           
           
$$\leftline{\indent$\displaystyle
 \sum_{n=1}^\infty {n^{n-3}\over  n!}(x e^{-x})^n  
= x - {3 x^2 \over 4}   +  {x^3 \over 6}   $}$$

\noindent Combinitorial identities and relations:
$$\leftline{\indent$\displaystyle  
\eqalign{
{ n-{1 \over 2} \choose n} 
   &= { \left(\textstyle{1 \over 2}\right)_n \over n!} 
    = {\Gamma(n+\textstyle{1 \over 2}) \over \Gamma({1 \over 2}) n! } 
   = { (2n-1)!! \over 2^{n} n! }
   = { (2n-1)! \over 2^{2n-1} n! (n-1)! }
   = { (2n)! \over 2^{2n} (n!)^2 }
    = {1 \over 2^{2n}} { 2n \choose n} \cr
   & = { 2 \over \pi } \int_0^{\pi \over 2} (\sin x)^{2n} dx  
     = { 2 \over \pi } \int_0^{\pi \over 2} (\cos x)^{2n} dx     \cr
   &\sim {1\over (\pi n)^{1/2} }  \exp\left(- {1 \over 8n} - {1 \over 192 n^3} + {1 \over 640 n^5}\ldots
           - {B_{2m}(2^{2m}-1) \over 2m(2m-1)2^{2m-1}n^{2m-1} }\ldots \right)  \cr
   &\sim {1\over (\pi n)^{1/2} }  \left[ 1 - {1 \over 8n} + {1 \over 128 n^2} - {1\over192 n^3} \ldots \right] }
     \hfill   {\bf 6.1.49}  $}$$

\centerline{\bf Stirling's Approximation}   
$$\leftline{\indent$\displaystyle
\ln \Gamma(n) = \ln (n-1)! \sim n \ln n - n - {1 \over 2} \ln n + {1 \over 2} \ln (2 \pi) + {1 \over 12 n}
- {1 \over 360 n^3} +  {1 \over 1260 n^5} -  {1 \over 1680 n^7} \ldots
    \hfill   {\bf 6.1.41}  $}$$                        

$$\leftline{\indent$\displaystyle
\ln n! \sim n \ln n - n + {1 \over 2} \ln n + {1 \over 2} \ln (2 \pi) + {1 \over 12 n}
- {1 \over 360 n^3} +  {1 \over 1260 n^5} -  {1 \over 1680 n^7} \ldots
 $}$$  
 
 $$\leftline{\indent$\displaystyle
 n! \approx \sqrt{2 \pi n} \left( {n \over e} \right)^n  \left(1 + {1 \over 12 n} + \ldots \right)
  $}$$                        
                                           
\bigskip
\centerline{\bf References}         
\parindent = 0pt

M. Abramowitz and I. Stegun, {\sl Handbook of Mathematical Functions},
Government Printing Office, also Dover, and on the web at
 http://www.math.sfu.ca/~cbm/aands/   [{\bf boldface} equation numbers above refer to formulas in this book].
Updated web version (2010), see http://dlmf.nist.gov/.

A. Erd\`elyi, et al., {\sl Higher Transcendental Functions}, McGraw Hill, 1953. 

I. S. Gradshteyn, I. M. Ryshik, {\sl Tables of Integrals, Series, Products}, Academic Press, 1980.

D. Knuth, {\sl The Art of Computer Programming, V. 1}, Addison-Wesley 1973, section 1.2.9.

H. S. Wilf, {\sl Generatingfunctionology} (Academic Press 2006).

\qquad 1994 edition: http://www.math.upenn.edu/\%7Ewilf/DownldGF.html
      
\end